\pdfoutput=1
\documentclass[conference]{IEEEtran}
\usepackage{cite}
\usepackage{amsmath,amssymb,amsfonts}
\usepackage{graphicx}
\usepackage{textcomp}
\usepackage{xcolor}
\usepackage{booktabs}
\usepackage{array}
\usepackage{pgfplots}
\usepackage{balance}
\usepackage[hyphens]{url}
\pgfplotsset{compat=1.17}
\usepgfplotslibrary{groupplots}
\newcommand{\Dext}{\ensuremath{D_{\mathrm{ext}}}}
\newcommand{\Mid}{\ensuremath{M}}

\begin{document}

\title{Hard Issue, Weak Model, or Bad Run?\\Rethinking Failure Prediction for Coding Agents}

\author{\IEEEauthorblockN{Xingyue Shi}
\IEEEauthorblockA{NC State University\\
Raleigh, NC, USA\\
xshi22@ncsu.edu}
\and
\IEEEauthorblockN{Wesley K. G. Assun\c{c}\~{a}o}
\IEEEauthorblockA{NC State University\\
Raleigh, NC, USA\\
wguezas@ncsu.edu}}

\maketitle

\begin{abstract}
Trajectory-based failure predictors for software engineering (SE) agents read the partial trajectory of a run to flag whether it is going right or wrong. 
If it is going wrong, the run can be stopped, retried, or switched to another model. We ask whether these predictors learn anything about the run itself. 
On public SWE-bench Verified trajectories, where each model attempts each issue once, two run-independent factors dominate predictor performance. 
The first is issue difficulty: a reference score that never looks at the run, only at how hard the issue is for other models, reaches 0.92 AUROC, whereas trajectory predictors reach about 0.62. 
The second is model ability: within an issue, knowing which model ran predicts failure better than our predictors do, and the predictors largely reproduce the overall ability ranking of the models. 
Beyond difficulty and ability, what predictors gain from the trajectory is statistically equivalent to zero. 
This confusion carries over to the actions taken after a prediction.
Model switching is often motivated by the gap between the best single model's resolution rate and an any-model oracle. 
This gap is no larger than random variation alone produces.
Per-issue switching gains nothing on held-out runs.
In a repeated-run pilot, the same model often solves an issue in some runs and fails it in others.
Retrying thus has real room to repair failures, but it rescues only failures caused by a bad run, not those on an issue beyond the model's ability.
Current predictors cannot tell the causes apart, as they mostly predict the issue and the model, not the run.
We call for failure prediction that explains why a run fails, evaluated on repeated-run benchmarks and in closed-loop, equal-budget settings.
\end{abstract}

\section{Introduction}
\label{sec:introduction}
Software engineering (SE) agents resolve repository issues by inspecting code, editing files, and running tests~\cite{jimenez2024swebench,swebenchverified}.
Agent runs are usually long and often fail~\cite{failfast,mehtiyev2026beyond}. 
A growing line of work therefore predicts failure while a run is still in progress and acts on the prediction: it stops the run to save tokens~\cite{agentstop,earlyeval}, restarts the same model~\cite{failfast}, or hands the issue to another model~\cite{atropos,swerouter}.
We focus on \emph{trajectory-based} predictors, which read only what the agent has produced so far (its reasoning, commands, and tool outputs) and therefore apply to closed frontier models.

A prediction can guide the choice of action only if it explains why the current run is likely to fail, using evidence from that run.
A run can fail for three different reasons.
The \emph{issue may be hard}, so that most models fail it;
the \emph{model may be weak}, so that it fails many issues;
or this particular attempt may have gone wrong, a \emph{bad run}, in which case the same model solves the same issue in other attempts.
This decomposition follows item response theory, which explains each outcome by the difficulty of the item and the ability of the respondent~\cite{rasch1960,rodriguez2021leaderboards,ge2026psychometrics}, and the run-to-run variation observed when the same agent repeats a task~\cite{yao2025taubench,monkeys,dtd}.
Each reason calls for a different action: a hard issue for stopping or human intervention, a weak model for switching to a stronger one, and only a bad run for retrying.

Existing work, however, rarely tells these three reasons apart.
We trace this to three assumptions that current practice takes for granted:
(i)~that a predictor with a high AUROC has learned run-specific information, that is, it predicts failure from the state of the current run;
(ii)~that the \emph{oracle gap}, the success rate of any model minus that of the best single model, reflects issue-specific strengths that switching can exploit~\cite{dei}; and
(iii)~that an alarm marks a bad run, so that acting on it, for example by retrying, pays off~\cite{failfast}.
We question whether these assumptions hold.
Standard evaluations cannot test them: predictors are usually scored with metrics pooled over many issues, and often over many models~\cite{failfast,agentstop,doomed,lph}.
Such a score mixes issue difficulty and model ability into one number, so a high value does not show that the predictor has learned anything from the run.
Even with repeated runs, the score is dominated by comparisons between different issues~\cite{dtd}.
That work separates issue difficulty from run-level failure within each model but does not examine model ability as a cause of failure or test the evidence for switching.
On public leaderboard trajectories, each model attempts each issue once, so a bad run cannot be told from a hard issue or a weak model.

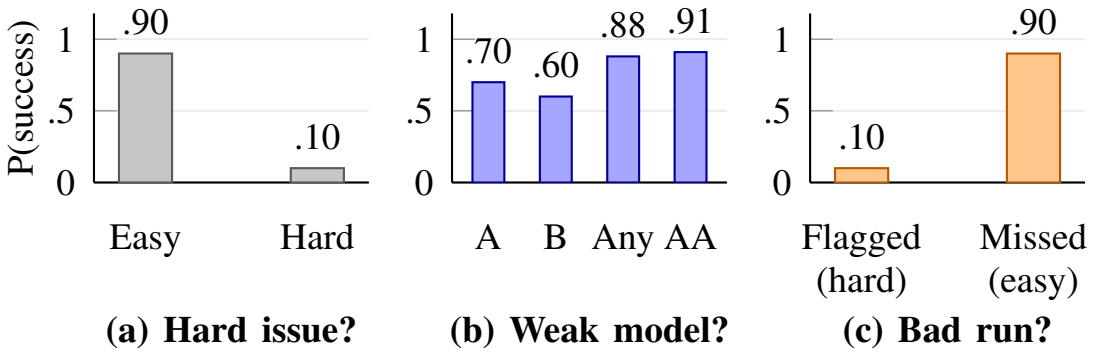
\begin{figure}[t]
\centering
\begin{tikzpicture}
\begin{groupplot}[group style={group size=3 by 1, horizontal sep=0.55cm},
  ybar, /pgf/bar width=10pt, width=3.4cm, height=2.7cm, ymin=0, ymax=1.18,
  ytick={0,0.5,1}, yticklabels={0,.5,1}, axis lines*=left, enlarge x limits=0.3,
  tick label style={font=\scriptsize}, xlabel style={font=\scriptsize\bfseries, at={(axis description cs:0.5,-0.68)}, anchor=north},,
  nodes near coords, every node near coord/.append style={font=\scriptsize},
  point meta=explicit symbolic,
  xtick style={draw=none}, ymajorgrids, grid style={gray!15}]
\nextgroupplot[xlabel={(a) Hard issue?}, xtick={1,2}, xticklabels={Easy,Hard}, ylabel={\scriptsize P(success)}, ylabel style={yshift=-1.2ex}]
  \addplot[fill=gray!45, draw=gray!70!black] coordinates {(1,0.9) [.90] (2,0.1) [.10]};
\nextgroupplot[xlabel={(b) Weak model?}, /pgf/bar width=6pt, enlarge x limits=0.18, xtick={1,2,3,4}, xticklabels={A,B,{Any},{AA}}]
  \addplot[fill=blue!35, draw=blue!60!black] coordinates {(1,0.7) [.70] (2,0.6) [.60] (3,0.88) [.88] (4,0.91) [.91]};
\nextgroupplot[xlabel={(c) Bad run?}, xtick={1,2}, xticklabels={{Flagged\\(hard)},{Missed\\(easy)}}, x tick label style={align=center}]
  \addplot[fill=orange!45, draw=orange!70!black] coordinates {(1,0.1) [.10] (2,0.9) [.90]};
\end{groupplot}
\end{tikzpicture}
\caption{A toy world in which each outcome depends only on the issue, the model, and a coin flip; values are probabilities, not empirical results. (a)~Success on easy and hard issues. (b)~Success with model A, model B, either model (Any), and A run twice (AA). (c)~Retry success after a failure that a difficulty score flags (hard issue) or misses (easy issue, a bad run).}

\label{fig:toy}
\end{figure}

Fig.~\ref{fig:toy} shows how all three assumptions can fail in a toy world where each run is a coin flip whose success probability is set only by the issue and the model.
There is thus nothing run-specific to learn, and no model is better than another on particular issues, yet the usual evidence for each assumption still appears.
(a)~\emph{A high pooled AUROC does not require learning about the run.}
Suppose a model succeeds with probability 0.9 on easy issues and 0.1 on hard ones, and the two kinds are equally common.
A score that only knows whether an issue is hard reaches 0.90 pooled AUROC, yet it cannot rank two runs of the same issue (0.50).
(b)~\emph{An oracle gap does not require issue-specific strengths.}
Suppose model~A succeeds with probability 0.7 and model~B with 0.6 on every issue.
Taking whichever model succeeds solves $1-(0.3\times0.4)=0.88$ of the issues, 18 percentage points (pp) more than A alone, although A is the better choice on every issue.
With the same two attempts, running A twice does even better (0.91): the gap comes from a second attempt, not from a better-suited model.
(c)~\emph{An alarm need not mark a bad run.}
In the world of (a), 90\% of failures occur on hard issues ($0.45$ of $0.50$), where a retry succeeds only 10\% of the time; failures on easy issues are bad runs, which a retry fixes 90\% of the time.
The difficulty score from (a) raises alarms on hard issues, so it flags the failures a retry rarely fixes and misses those it usually fixes.

We ask whether real agents behave like this toy world, turning each assumption into a research question:
\textbf{RQ1:} Do failure predictors learn about the run, beyond issue difficulty and model ability?
\textbf{RQ2:} Does the oracle gap reflect issue-specific strengths that switching can exploit?
\textbf{RQ3:} Do alarms mark bad runs that a retry can fix?
On trajectories of seven recent models, outcomes of 23 models, and a public corpus with repeated runs, we find that real agents largely behave like the toy world, and the data support none of the three assumptions.
Beyond issue difficulty and model ability, what predictors gain from the trajectory is statistically equivalent to zero (RQ1).
The oracle gap is no larger than chance alone produces, and choosing a model for each issue from past runs brings no measurable gain on new runs (RQ2).
Bad runs are common enough for retrying to pay off, but current predictors can hardly target them (RQ3).

Rather than proposing another predictor, we question what current predictors predict.
Our answer is that they mostly predict the issue and the model, not the run, and that the evidence for acting on their alarms, whether by switching or by retrying, inherits the same confusion.
We show how each assumption can be tested with existing data (Section~\ref{sec:new-light}) and call for failure prediction that explains why a run fails, so that each failure is met with the action that can fix it.

\section{Methodology}
\label{sec:setting}
\textbf{Data.}
To study predictors where they are usually built, we use public leaderboard-style trajectories: mini-swe-agent runs on the 500 issues of SWE-bench Verified, in which each model attempts each issue once~\cite{miniswe,ds_traj}.
They include 3,492 runs with known outcomes from seven models (Claude Haiku~4.5, Claude Opus~4.5 and~4.6, Gemini~3 Flash, GLM-5, GPT-5-mini, and MiniMax-M2.5), 994 of which fail.
We use the released resolved labels as failure labels and exclude two GPT-5.2 sources that contain identical transcripts with conflicting labels.
A companion release covers 25 models on the same issues~\cite{ds_25}. 
After the same exclusion, it gives the outcomes of 23 models (the 23-model data), to test whether the findings hold for a larger, more diverse set of models, and 16 models outside the seven, used only to estimate issue difficulty.

\textbf{Predictors.}
Our predictors mirror the design of Fail-Fast~\cite{failfast}.
A predictor reads a run's trajectory after $t\in\{10,20,40\}$ completed steps, either the last eight steps (R8) or the full prefix (ALL), and outputs a failure score.
Earlier observation points leave more runs to act on but give less evidence, and the two windows contrast recent context with the full history.
Unlike Fail-Fast, a predictor in this work sees neither the issue text nor the name of the model, so whatever it learns comes from the trajectory alone, which may still reveal the issue and the model implicitly.
We train TF-IDF logistic regression, a choice also used in~\cite{dtd}, with issue-grouped five-fold cross-validation. 
To check that the findings are not specific to a simple text model, we also fine-tune Qwen3-0.6B with LoRA on one fold, and pair each neural predictor with a logistic model on the same truncated text.
This yields 18 configurations: ten TF-IDF variants, covering the observation points, the windows, and the run subsets used to compare observation points, and four neural predictors with their paired logistic models. 
All configurations enter the tests, and Table~\ref{tab:pairs} shows two representative ones.
Unless noted, we report the 2,840 runs still running at step 20 (t20), since a predictor can act only on such runs.

\textbf{Run-blind references.}
To measure how much of a predictor's performance requires the run at all, we compare it with scores that ignore what the agent did.
\emph{Issue difficulty} (\Dext) is the fraction of the 16 outside models that fail the issue, so it assigns every run of an issue the same score; since it relies on outcomes unavailable at deployment, it serves as a diagnostic rather than a practical predictor.
\emph{Model identity} (\Mid) scores a run by its model's failure rate on training issues, that is, by the model's overall ability.
\emph{Length + \Mid} adds only the prefix length to \Mid, testing whether a predictor learns more than how long the run has lasted.

\textbf{Pairs.}
To see where a predictor's AUROC comes from, we decompose it by pair: AUROC is the fraction of failed--successful pairs that a predictor ranks correctly, that is, in which it scores the failed run higher.
We split these pairs into \emph{cross-issue} pairs and \emph{within-issue} pairs (same issue, different models), and call a within-issue pair \emph{consistent} if the failed run comes from the overall weaker model and an \emph{upset} otherwise.
A predictor that tracks the run should rank both kinds of within-issue pairs above chance, whereas a predictor that merely follows the ability ranking of the models ranks consistent pairs correctly and upsets incorrectly.

\textbf{Tests.} Three of the claims rely on dedicated tests.
To claim that a predictor adds nothing beyond \Dext\ and \Mid, we require the 90\% interval of its AUROC gain to lie within a pre-specified $\pm0.02$ (equivalence test, Holm-corrected over 50 comparisons~\cite{schuirmann1987,holm1979}).
To claim that it cannot tell which runs of an issue failed, we use a conditional logit~\cite{chamberlain1980}, which holds issue difficulty fixed and tests whether the score adds to \Mid.
To claim that the oracle gap (issues solved by some model but not the best one) and the upset rate, often read as evidence of issue-specific strengths, do not exceed chance, we build a \emph{no-specialization reference}.
It simulates outcomes in a world that matches the real models' abilities and the issues' difficulties, but in which every issue ranks the models identically, so that upsets and the gap arise only by chance.
We fit this world with the Rasch model~\cite{rasch1960}, which gives each model one ability and each issue one difficulty, and with the two-parameter logistic (2PL) model~\cite{birnbaum1968}, which also lets some issues separate the models more sharply and so leaves less to chance.

\textbf{Repeated-run pilot.}
Single-run data cannot tell specialization from luck, nor show what a retry brings.
We therefore use a public SWE-agent corpus with repeated runs~\cite{nebius}: two Llama models (70B and 8B), each run at least four times on each of 522 issues from SWE-bench-extra and SWE-bench dev.
To judge a choice on runs it has not seen, we split the runs of each model on each issue into two halves: half A is used to choose a model for each issue or to find upsets, and half B to measure success or check whether the upsets recur.
Because a single split is arbitrary, we average over 200 random splits, and we bootstrap issues, repeating the choice in every resample, to obtain intervals.

\section{Preliminary Findings}

\begin{table}[t]
\centering
\caption{AUROC by type of pair, on 2,840 runs still running at step 20.}
\label{tab:pairs}
\footnotesize
\setlength{\tabcolsep}{3.2pt}
\begin{tabular}{@{}lcccc@{}}
\toprule
 & & \multicolumn{3}{c}{Within-issue pairs} \\
\cmidrule(l){3-5}
 & Pooled & All & Consistent & Upset \\
\midrule
\multicolumn{5}{@{}l}{\emph{Run-blind references}} \\
\quad Issue difficulty (\Dext) & 0.924 & 0.500 & 0.500 & 0.500 \\
\quad Model identity (\Mid) & 0.567 & 0.724 & 1.000 & 0.000 \\
\quad \Dext\ + \Mid & 0.929 & 0.718 & 0.869 & 0.317 \\
\quad Length + \Mid & 0.607 & 0.648 & 0.756 & 0.373 \\
\midrule
\multicolumn{5}{@{}l}{\emph{Trajectory-based predictors}} \\
\quad TF-IDF, last 8 steps (R8) & 0.624 & 0.604 & 0.678 & 0.415 \\
\quad TF-IDF, full prefix (ALL) & 0.630 & 0.610 & 0.687 & 0.416 \\
\bottomrule
\end{tabular}
\end{table}

\subsection{Predictors Learn Mostly the Issue and the Model (RQ1)}
A high AUROC is taken to mean that a predictor has learned run-specific information, yet the toy world in Fig.~\ref{fig:toy}(a) shows that a score can reach a high AUROC without learning anything about the run.
We therefore compare the predictors with run-blind references for issue difficulty and model ability (Table~\ref{tab:pairs}) and test what they add beyond both.

\textbf{Issue difficulty explains pooled AUROC.}
Pooled AUROC is almost entirely a cross-issue measure.
Of the 494,109 failed--successful pairs in the test folds, 99.66\% compare different issues, as also observed on repeated-run benchmarks~\cite{dtd}.
A score that recognizes hard issues can therefore score highly without seeing the run.
The difficulty score \Dext\ reaches 0.92 pooled AUROC, whereas the predictors reach only 0.62--0.63 and are not significantly better than length plus \Mid\ (all ten 95\% intervals of the difference contain 0).

\textbf{Model ability explains within-issue AUROC.}
Within an issue, the question becomes which model's run failed, and knowing the model answers it better than the trajectory-based predictors do.
Model identity \Mid\ reaches 0.72 within-issue AUROC, above both predictors (0.60--0.61), and 0.82 on the 23-model data.
By construction, \Dext\ scores 0.5 within an issue, and \Mid\ scores 1 on consistent pairs and 0 on upsets.
The predictors are never told the model, yet they show the same pattern in weaker form.
They score 0.68--0.69 on consistent pairs but 0.42 on upsets, and all ten distinct TF-IDF configurations fall below 0.5 on upsets.
A predictor that follows the run should instead exceed 0.5 on both.
Ability matters more when models differ more, explaining 2.5\% of outcome variance with seven models but about 10\% with 23.

\textbf{The trajectory adds nothing measurable beyond both.}
Adding a predictor's score to \Dext\ and \Mid\ changes pooled AUROC by an amount equivalent to zero in all 18 configurations (e.g., $-0.001$ for the full-prefix predictor in Table~\ref{tab:pairs}, 90\% interval [$-0.003$, 0.000]).
Within issues, a conditional logit that holds issue difficulty fixed shows that adding the score to \Mid\ does not help identify which runs of an issue failed in any configuration (e.g., $-0.011$ within-issue AUROC for the same predictor, 95\% interval [$-0.024$, 0.002]).

\begin{figure}[t]
\centering
{\scriptsize
\tikz[baseline=-0.5ex]{\draw[|-|,thick,gray] (0,0)--(0.35,0);} Rasch
\quad \tikz[baseline=-0.5ex]{\draw[|-|,thick,blue!70!black] (0,0)--(0.35,0);} 2PL 
\quad \tikz[baseline=-0.5ex]{\fill[red] (0,0) circle (1.6pt);} Observed\par}
\smallskip
\begin{tikzpicture}
\begin{groupplot}[group style={group size=2 by 1, horizontal sep=0.3cm, yticklabels at=edge left},
  scale only axis, width=3.05cm, height=1.75cm, ymin=0.45, ymax=2.55,
  ytick={1,2}, yticklabels={23 models, 7 models},
  tick label style={font=\scriptsize}, label style={font=\scriptsize}, xlabel style={font=\scriptsize\bfseries},
  xmajorgrids, grid style={gray!20}, ytick style={draw=none}]
\nextgroupplot[xlabel={(a) Upset rate (\%)}, xmin=15, xmax=30.5, xtick={15,20,25,30}]
  \draw[|-|, thick, gray] (axis cs:22.99,2.17) -- (axis cs:29.17,2.17);
  \draw[|-|, thick, blue!70!black] (axis cs:20.26,1.83) -- (axis cs:26.26,1.83);
  \draw[|-|, thick, gray] (axis cs:18.93,1.17) -- (axis cs:21.19,1.17);
  \draw[|-|, thick, blue!70!black] (axis cs:16.24,0.83) -- (axis cs:18.23,0.83);
  \draw[red, thick] (axis cs:24.64,1.62) -- (axis cs:24.64,2.38);
  \draw[red, thick] (axis cs:17.74,0.62) -- (axis cs:17.74,1.38);
  \addplot[only marks, mark=*, mark size=1.6pt, red] coordinates {(24.64,2) (17.74,1)};
\nextgroupplot[xlabel={(b) Oracle gap (pp)}, xmin=7, xmax=20, xtick={8,12,16,20}]
  \draw[|-|, thick, gray] (axis cs:8.70,2.17) -- (axis cs:12.40,2.17);
  \draw[|-|, thick, blue!70!black] (axis cs:8.20,1.83) -- (axis cs:12.40,1.83);
  \draw[|-|, thick, gray] (axis cs:14.80,1.17) -- (axis cs:19.20,1.17);
  \draw[|-|, thick, blue!70!black] (axis cs:11.40,0.83) -- (axis cs:15.20,0.83);
  \draw[red, thick] (axis cs:10.0,1.62) -- (axis cs:10.0,2.38);
  \draw[red, thick] (axis cs:12.4,0.62) -- (axis cs:12.4,1.38);
  \addplot[only marks, mark=*, mark size=1.6pt, red] coordinates {(10.0,2) (12.4,1)};
\end{groupplot}
\end{tikzpicture}
\caption{Observed values (red) versus the no-specialization reference, simulated with Rasch and 2PL models in which every issue ranks the models identically. Bars show the middle 95\% of 500 simulations. (a)~Upset rate, the share of one-success model pairs won by the weaker model. (b)~Oracle gap, the resolution rate of any model minus that of the best single model.}

\label{fig:null}
\end{figure}
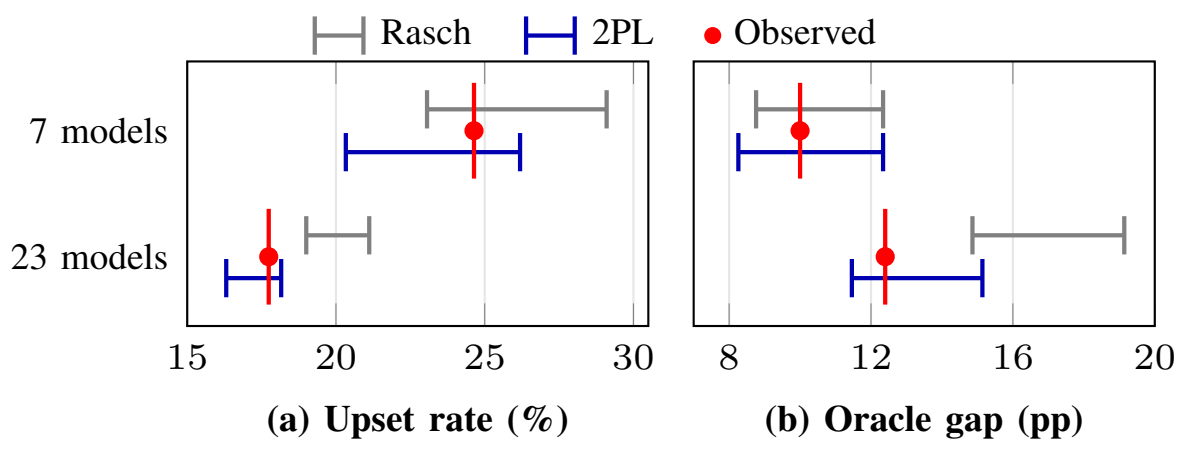

\subsection{The Oracle Gap Is Not a Gain from Switching (RQ2)}
Switching presumes that a model failing an issue can be rescued by another model that is better on it, and the usual evidence is the oracle gap.
Fig.~\ref{fig:toy}(b) shows that such a gap can arise without issue-specific strengths; we test whether the real one does, and whether switching can turn it into a gain.

\textbf{The oracle gap is not evidence of specialization.}
The best single model solves 76.8\% of the issues, whereas some model solves 86.8\%, an oracle gap of 10.0~pp (12.4~pp with 23 models).
Upsets are also common.
When exactly one of two models solves an issue, the overall weaker model succeeds in 24.6\% of cases (17.7\% with 23 models).
Fig.~\ref{fig:null} compares both quantities with the range that chance alone produces in the no-specialization reference.
For 7 and 23 models, the observed values lie within the 2PL range, and with 23 models the Rasch range lies even above them.
In none of the eight comparisons (oracle gap and upset rate, 7 and 23 models, Rasch and 2PL) is an observed value significantly above its reference after Holm correction.
The gap is therefore no evidence of specialization, though this does not show that specialization is absent.

\textbf{The oracle gap is not an attainable gain.}
For 45\% of all failures (448 of 994) there is no model to switch to, since they occur on the 64 issues that none of the seven models solves.
Moreover, the oracle picks a model after seeing the outcomes, whereas a router must pick before.
Choosing before the run gains little, even with the issue's own history.
In the repeated-run pilot, choosing for each issue the model that did better on its past runs (half A) gains only $+0.09$~pp on new runs (half B) over always using the model that is better overall (Table~\ref{tab:actions}).
It gains little on new issues either.
A router trained on other issues predicts each model's chance of solving a new issue from the model's identity and the issue text, before any run, and picks the most promising model, yet it gains only $+0.41$~pp over the best single model, the one with the highest resolution rate on the training issues.

\textbf{Specialization is real but sparse.}
If upsets were pure chance, an upset found in half A of the runs in the repeated-run pilot would recur in half B only at the rate the no-specialization reference predicts (19\%).
Instead, 44\% recur ($p=0.004$), indicating that some models are better on particular issues.
These recurring upsets, however, occur on only about 15 of the 522 issues.
We decompose the variation in outcomes across repeated runs into parts due to the issue, the model, their combination (specialization), and run-to-run variation.
Specialization accounts for only 4.6\%, against 53\% for issue difficulty and 42\% for run-to-run variation.

\textbf{Escalation gains come from the extra attempt.}
On the seven-model data, escalating every failure to one fixed second model gains $+6.91$~pp (Table~\ref{tab:actions}), and no other pair of models chosen in hindsight does better.
Because this policy makes no issue-specific choice, its gain comes from the second attempt, not from picking a better-suited model; choosing a model per issue alone adds only $+0.41$~pp.
The gain also assumes a perfect failure detector.
RQ3 further asks what a second attempt with the same model can fix.

\begin{table}[t]
\centering
\caption{Resolution rate of strategy A minus that of B (pp).}
\label{tab:actions}
\footnotesize
\setlength{\tabcolsep}{3pt}
\begin{tabular}{@{}lc@{}}
\toprule
A vs.\ B & A $-$ B [95\% CI] \\
\midrule
\multicolumn{2}{@{}l}{\emph{One attempt: choose a model per issue?}} \\
\quad Choice from the issue's past runs vs.\ one model & $+0.09$ [$-0.36$, 0.63] \\
\quad Router for new issues vs.\ best single model & $+0.41$ [$-1.83$, 2.85] \\
\midrule
\multicolumn{2}{@{}l}{\emph{One vs.\ two attempts: escalate after failure?}} \\
\quad Escalation vs.\ best single model & $+6.91$ [\phantom{$-$}4.67, 9.15] \\
\bottomrule
\end{tabular}
\end{table}

\subsection{Alarms Miss the Runs a Retry Can Fix (RQ3)}
Retrying after an alarm assumes that the failure is a bad run, that is, that the model can solve the issue and only this attempt went wrong.
If instead the issue is hard or the model is weak, a second run will likely fail again, and stopping, switching to a stronger model, or asking a human is the better action.
An alarm can therefore guide a retry only if it tells a bad run from a hard issue or a weak model.
In Fig.~\ref{fig:toy}(c), a difficulty score flags mostly hard issues; we test whether real alarms do too.

\textbf{Bad runs are common.}
If failures came only from hard issues and weak models, a model would fail an issue on every run.
In the repeated-run pilot, this is not the case.
The same model often solves an issue in some runs and fails it in others; this run-to-run variation accounts for 42\% of the variation in outcomes, far more than specialization (4.6\%; RQ2).
Each such failure is a bad run, since the same model solves the issue in other attempts, so a retry has a real chance to fix it.

\textbf{Alarms cannot tell a bad run from a hard issue or a weak model.}
An alarm should not only flag a likely failure but also indicate whether a retry can fix it.
The predictors' scores, however, carry nothing measurable beyond issue difficulty and model identity (RQ1), and 64--82\% of the variance of their scores lies between issues rather than within them.
Much like the difficulty score in Fig.~\ref{fig:toy}(c), a high score thus signals a hard issue or a weak model, where a retry is less likely to succeed, and says little about whether this run went wrong.
This is consistent with Fail-Fast~\cite{failfast}, in which a cold restart triggered by its predictor adds only 0.2~pp.

\section{Discussion}
\label{sec:new-light}
\textbf{Rethinking failure prediction.}
Our findings challenge three assumptions that current work on failure prediction for coding agents takes for granted.
A high AUROC, a large oracle gap, and an alarm can each look like evidence that prediction helps, yet each can arise from issue difficulty, model ability, or chance alone.
What these assumptions overlook is the \emph{cause} of a failure, which decides whether stopping, switching, or retrying can help.
Future work can evaluate predictors more rigorously with three tests designed in this work that need only existing data.
First, report within-issue AUROC next to pooled AUROC and compare it with run-blind references for issue difficulty, model ability, and length.
Second, compare any oracle gap with a no-specialization reference and evaluate routers on held-out runs.
Third, judge an alarm by whether the action it triggers, such as a retry, fixes the failure, and compare actions at equal numbers of attempts.

\textbf{Threats to validity.}
The evidence comes mainly from released labels on one benchmark and scaffold, from trajectory-based predictors only, and partly from the repeated-run pilot with older models and a router with shallow issue features.
We have not yet measured how much specialization the no-specialization reference can detect, so non-rejection does not prove that specialization is absent.

\section{Future Plans}
\label{sec:future}

These findings suggest that failure prediction should target the cause of a failure, not only the failure itself.
We plan three steps toward this goal.
First, we will collect repeated runs of recent models across more benchmarks and scaffolds, so that run-to-run variation can be separated from issue difficulty and model ability.
Second, we will build predictors that tell why a run fails, because each cause calls for a different action.
They will separate a hard issue, a weak model, and a bad run, so that an alarm can trigger stopping, switching, or retrying accordingly.
Third, we will evaluate predictors in a closed loop at equal budgets, counting false alarms and the cost of monitoring, and judge each by the decisions it improves rather than by its pooled AUROC.

\balance
\bibliographystyle{IEEEtran}
\bibliography{references}

\begin{thebibliography}{10}
\providecommand{\url}[1]{#1}
\csname url@samestyle\endcsname
\providecommand{\newblock}{\relax}
\providecommand{\bibinfo}[2]{#2}
\providecommand{\BIBentrySTDinterwordspacing}{\spaceskip=0pt\relax}
\providecommand{\BIBentryALTinterwordstretchfactor}{4}
\providecommand{\BIBentryALTinterwordspacing}{\spaceskip=\fontdimen2\font plus
\BIBentryALTinterwordstretchfactor\fontdimen3\font minus
  \fontdimen4\font\relax}
\providecommand{\BIBforeignlanguage}[2]{{%
\expandafter\ifx\csname l@#1\endcsname\relax
\typeout{** WARNING: IEEEtran.bst: No hyphenation pattern has been}%
\typeout{** loaded for the language `#1'. Using the pattern for}%
\typeout{** the default language instead.}%
\else
\language=\csname l@#1\endcsname
\fi
#2}}
\providecommand{\BIBdecl}{\relax}
\BIBdecl

\bibitem{jimenez2024swebench}
C.~E. Jimenez, J.~Yang, A.~Wettig, S.~Yao, K.~Pei, O.~Press, and K.~Narasimhan,
  ``{SWE}-bench: {Can} language models resolve real-world {GithHub} issues?''
  in \emph{International {Conference} on {Learning} {Representations}}, 2024,
  pp. 54\,107--54\,157.

\bibitem{swebenchverified}
{OpenAI}, ``Introducing {SWE-bench} verified,''
  \url{https://openai.com/index/introducing-swe-bench-verified/}, 2024,
  accessed: 2026.

\bibitem{failfast}
C.~Wang, Y.~Lyu, J.~He, Z.~Yang, C.~Zhong, Y.~Harel, and D.~Lo, ``{Fail-Fast},
  {Restart-Smart}: Early failure prediction and restart for {SWE} agentic
  tasks,'' \emph{arXiv preprint arXiv:2608.03222}, 2026.

\bibitem{mehtiyev2026beyond}
T.~Mehtiyev and W.~Assun{\c{c}}{\~a}o, ``Beyond resolution rates: Behavioral
  drivers of coding agent success and failure,'' \emph{arXiv preprint
  arXiv:2604.02547}, 2026.

\bibitem{agentstop}
D.~Pham, K.~Katevas, A.~S. Shamsabadi, and H.~Haddadi, ``{AgentStop}:
  Terminating local {AI} agents early to save energy in consumer devices,'' in
  \emph{Proceedings of the ACM Conference on AI and Agentic Systems}.\hskip 1em
  plus 0.5em minus 0.4em\relax Association for Computing Machinery, 2026, pp.
  1051--1069.

\bibitem{earlyeval}
Y.~Shi, Z.~Sun, J.~Dong, C.~Wan, D.~Lo, and X.~Gu, ``{EarlyEval}: Cheaper agent
  evaluation via early outcome prediction,'' \emph{arXiv preprint
  arXiv:2609.02783}, 2026.

\bibitem{atropos}
N.~Kim and S.~Yoo, ``Atropos: Improving cost-benefit trade-off of {LLM}-based
  agents under self-consistency with early termination and model hotswap,''
  \emph{Proceedings of the ACM on Software Engineering}, vol.~3, no. ISSTA, pp.
  1492--1512, 2026.

\bibitem{swerouter}
S.~Son, S.~Yoon, J.~Tang, S.~Wang, L.~Wolf, and I.~Bogunovic, ``{SWE-Router}:
  Routing in multi-turn agentic software engineering tasks,'' \emph{arXiv
  preprint arXiv:2607.00053}, 2026.

\bibitem{rasch1960}
G.~Rasch, \emph{Probabilistic models for some intelligence and attainment
  tests.}\hskip 1em plus 0.5em minus 0.4em\relax Chicago: MESA Press, 1993.

\bibitem{rodriguez2021leaderboards}
P.~Rodriguez, J.~Barrow, A.~Hoyle, J.~P. Lalor, R.~Jia, and J.~Boyd-Graber,
  ``Evaluation examples are not equally informative: How should that change
  {NLP} leaderboards?'' in \emph{Proceedings of the 59th Annual Meeting of the
  {A}ssociation for {C}omputational {L}inguistics and the 11th International
  Joint Conference on Natural Language Processing (Volume 1: Long Papers)},
  2021, pp. 4486--4503.

\bibitem{ge2026psychometrics}
C.~Ge, D.~Kryvosheieva, D.~Fried, U.~Girit, and K.~Hariharan, ``Agent
  psychometrics: Task-level performance prediction in agentic coding
  benchmarks,'' \emph{arXiv preprint arXiv:2604.00594}, 2026.

\bibitem{yao2025taubench}
S.~Yao, N.~Shinn, P.~Razavi, and K.~Narasimhan, ``$\tau$-bench: A benchmark for
  tool-agent-user interaction in real-world domains,'' in \emph{The Thirteenth
  International Conference on Learning Representations}, 2025, pp.
  74\,824--74\,876.

\bibitem{monkeys}
B.~Brown, J.~Juravsky, R.~Ehrlich, R.~Clark, Q.~V. Le, C.~R{\'e}, and
  A.~Mirhoseini, ``Large language monkeys: Scaling inference compute with
  repeated sampling,'' \emph{arXiv preprint arXiv:2407.21787}, 2024.

\bibitem{dtd}
M.~EsfandyariDoulabi, L.~Arkoh, B.~Tadesse, V.~Patel, M.~Sharma, M.~d'Amorim,
  and W.~Assun{\c{c}}{\~a}o, ``Disentangling task difficulty from run-level
  failure in agent failure prediction,'' \emph{arXiv preprint
  arXiv:2610.05572}, 2026.

\bibitem{dei}
K.~Zhang, W.~Yao, Z.~Liu, Y.~Feng, Z.~Liu, R.~R.~N. Murthy, T.~Lan, L.~Li,
  R.~Lou, J.~Xu, B.~Pang, Y.~Zhou, S.~Heinecke, S.~Savarese, H.~Wang, and
  C.~Xiong, ``Diversity empowers intelligence: Integrating expertise of
  software engineering agents,'' in \emph{The Thirteenth International
  Conference on Learning Representations}, 2025, pp. 86\,691--86\,714.

\bibitem{doomed}
K.~Ruan, Z.~Huang, Z.~Zhou, Q.~Wei, J.~Lin, X.~Wang, and H.~Sun, ``Doomed from
  the start: Early abort of {LLM} agent episodes via a recall-controlled probe
  cascade,'' \emph{arXiv preprint arXiv:2607.06503}, 2026.

\bibitem{lph}
A.~Silva, H.~Tu, and M.~Monperrus, ``Latent programming horizons in coding
  agents,'' \emph{arXiv preprint arXiv:2607.05188}, 2026.

\bibitem{miniswe}
{SWE-agent contributors}, ``{mini-swe-agent},''
  \url{https://github.com/SWE-agent/mini-swe-agent}, accessed: 2026.

\bibitem{ds_traj}
{tarsur385}, ``swebench-verified-trajectories,''
  \url{https://huggingface.co/datasets/tarsur385/swebench-verified-trajectories},
  accessed: 2026.

\bibitem{ds_25}
------, ``swev-trm-trajectories-25models,''
  \url{https://huggingface.co/datasets/tarsur385/swev-trm-trajectories-25models},
  accessed: 2026.

\bibitem{schuirmann1987}
D.~J. Schuirmann, ``A comparison of the two one-sided tests procedure and the
  power approach for assessing the equivalence of average bioavailability,''
  \emph{Journal of Pharmacokinetics and Biopharmaceutics}, vol.~15, no.~6, pp.
  657--680, 1987.

\bibitem{holm1979}
S.~Holm, ``A simple sequentially rejective multiple test procedure,''
  \emph{Scandinavian Journal of Statistics}, pp. 65--70, 1979.

\bibitem{chamberlain1980}
G.~Chamberlain, ``Analysis of covariance with qualitative data,'' \emph{The
  Review of Economic Studies}, vol.~47, no.~1, pp. 225--238, 1980.

\bibitem{birnbaum1968}
A.~Birnbaum, ``Some latent trait models and their use in inferring an
  examinee's ability,'' \emph{Statistical Theories of Mental Test Scores},
  1968.

\bibitem{nebius}
{Nebius}, ``{SWE}-agent-trajectories,''
  \url{https://huggingface.co/datasets/nebius/SWE-agent-trajectories},
  accessed: 2026.

\end{thebibliography}

\end{document}